\pdfoutput=1

\documentclass[letterpaper]{article} 
\usepackage{aaai23-r2hcai}  
\usepackage{fancyhdr}
\usepackage{graphicx}
\usepackage[english]{babel}
\usepackage[utf8]{inputenc}
\usepackage[T1]{fontenc}
\usepackage{lmodern}
\usepackage{caption}

\usepackage[
    citestyle=numeric,
    style=numeric,
    backend=biber,
    sorting=none
    ]{biblatex}
\usepackage{csquotes}
\usepackage{braket}
\usepackage{amsmath}
\usepackage{amssymb}
\usepackage{physics}
\usepackage{svg}
\usepackage{tikz}
\usepackage{quantikz}
\usepackage{booktabs}
\usepackage{tcolorbox}
\usepackage{amsthm}
\usepackage{txfonts}
\usepackage{mathtools}
\usepackage[raggedright]{titlesec}
\usepackage{csquotes}
\usepackage{esvect}

\usepackage{algorithm2e}

\usepackage{newfloat}
\usepackage{listings}

\DeclareCaptionStyle{ruled}{labelfont=normalfont,labelsep=colon,strut=off} 
 
\usepackage{fancyhdr}
\title{QNEAT: Natural Evolution of Variational Quantum Circuit Architecture}
\author{
    Alessandro Giovagnoli $^1$ \quad
    Yunpu Ma $^2$ \quad
    Volker Tresp $^3$
}
\affiliations {
    Ludwig Maximilians University, Munich, Germany \quad
    \\
    \textsuperscript{\rm 1} a.giovagnoli@campus.lmu.de, \textsuperscript{\rm 2} 
 cognitive.yunpu@gmail.com, \textsuperscript{\rm 3} volker.tresp@lmu.de
}

\begin{document}
\maketitle

\begin{abstract}
Quantum Machine Learning (QML) is a recent and rapidly evolving field where the theoretical framework and logic of quantum mechanics are employed to solve machine learning tasks. Various techniques with different levels of quantum-classical hybridization have been proposed. Here we focus on variational quantum circuits (VQC), which emerged as the most promising candidates for the quantum counterpart of neural networks in the noisy intermediate-scale quantum (NISQ) era. Although showing promising results, VQCs can be hard to train because of different issues, e.g., barren plateau, periodicity of the weights, or choice of architecture. This paper focuses on this last problem for finding optimal architectures of variational quantum circuits for various tasks. To address it, we propose a gradient-free algorithm inspired by natural evolution to optimize both the weights and the architecture of the VQC. In particular, we present a version of the well-known neuroevolution of augmenting topologies (NEAT) algorithm and adapt it to the case of variational quantum circuits. We refer to the proposed architecture search algorithm for VQC as QNEAT. We test the algorithm with different benchmark problems of classical fields of machine learning i.e. reinforcement learning and combinatorial optimization.
\end{abstract}

\section{Introduction}

    The field of quantum computing has, in the last decades, attracted much attention due to the promise of enabling us to solve problems that classically, altaugh they be solved, are practically not feasible \cite{better1}, \cite{better2}, \cite{better3}, \cite{better4}. These problems range from quantum metrology \cite{metrology1}, \cite{metrology2}, mathematics \cite{math1}, \cite{math2}, chemistry \cite{chem1} to optimization in general \cite{optim1}, \cite{optim2}, \cite{optim3}. The possibility of using intrinsic quantum algorithms that proved to have a speedup like the Deutsch–Jozsa, Grover search, quantum Fourier transform, or a hybrid of classical and quantum techniques \cite{hybrid1} is pushing the academy and industry into exploring the extent of the capabilities of this field. 

    At this stage, the theoretical methods and algorithms related to the field of quantum information theory are limited by the current state of the art in quantum hardware production, which is developing at a slower pace.
    
    In the current noisy intermediate-scale quantum (NISQ) era, quantum computers contain qubits ranging from 50 to 100, which are not fault-tolerant. They are still affected by decoherence, meaning that the qubits cannot be kept for a long time in the desired quantum state, and are not able to continuously implement quantum error correction, features needed to achieve the so-called quantum supremacy \cite{prob1}, \cite{prob2}, \cite{prob2}.

    In the current NISQ era emerged the field of quantum machine learning (QML), which exploits and mixes classical and quantum techniques to solve classical or pure quantum machine learning problems \cite{qml}. Some algorithms and techniques have already proved to give a quantum speedup over their classical counterparts once the hardware allows managing a higher number of qubits with higher stability. Other methods are instead being still explored. 
    
    A promising class of algorithms is the variational quantum algorithms that use variational quantum circuits (VQCs) as the building block. In these parametrized quantum circuits, the physical gates, usually rotations and controlled-nots, depend on adjustable parameters that can be adapted to make the circuit perform better on various tasks. 
  
    Similar to classical machine learning, we use neural networks as function approximators that rely on the theory of the well-known Universal Approximation Theorem \cite{universalapprox}. Analogously we assume here the existence of a function mapping the quantum state containing the features of the problem of interest into the quantum state containing the labels. According to the laws of quantum mechanics, this function is a unitary evolution mapping an initial state into the final one. It can be shown that any unitary operator acting on multiple qubits can be expressed through ROTs and CNOTs, which justifies the interest in using VQCs and exploring their capabilities to solve machine-learning tasks.

    Due to the low amount of qubits needed to encode the information and the low amount of parameters necessary to approximate the function of interest adequately, VQCs are promising candidates for quantum neural networks in the NISQ era. They are already employed in several machine learning tasks. In these tasks, VQCs are set in a classical neural network pipeline, and their weights are trained to minimize a selected loss function once a fixed architecture has been chosen. 
    
    Nonetheless, they present issues that may limit their usage. One major problem is the classical barren plateau phenomenon \cite{bp1}, \cite{bp2}, \cite{bp3}, \cite{bp4}, \cite{bp5}, which is linked to the expressibility of VQCs and not only to the optimizer. Another issue is that the depth of the circuit increases the noise, leading to the loss of entanglement or coherence between qubits. A key factor in the study of the VQCs is thus choosing an appropriate architecture, or \textit{ansatz}, that can adequately approximate the function of interest. However, simultaneously, it doesn't compromise the stability of the circuit by making it too deep, i.e., with an unnecessary amount of quantum gates. This motivates the interest in studying variable architecture algorithms that evolve, finding the most suitable architecture and weights rather than starting from a fixed architecture and only updating its weights.  

    In Section~\ref{sec:relatedwork} we will first briefly review some of the techniques already shown in the literature to perform such tasks. In Section~\ref{sec:qneat} and \ref{sec:algorithm} we present our algorithm, in Section~\ref{sec:experiments} the experiments setting on which it has been tested on and in Section~\ref{sec:results} we present the results. In conclusion, in Section~\ref{sec:MOO} we show a variation of the QNEAT algorithm for multiple objective optimization purposes.

\section{Related work\label{sec:relatedwork}}
    The classical machine learning pipeline, which has also been applied to VQCs, is based on the traditional gradient-based approach. In this approach, a selected loss function is evaluated, gradients are computed through the backpropagation, and the weights of the function approximator are updated. 
    
    One of the main problems in applying the classical gradient-based methods is the barren plateau of gradients during the training of VQCs. It has been shown \cite{bp6} that with the increasing depth of the circuit, the probability of finding a non-zero entry in the gradient becomes exponentially small. This issue is related to the intrinsic expressibility of variational quantum circuits and not simply to the selected optimizer. Therefore, gradient-free methods have been proposed to overcome problems, such as barren plateaus or the risk of being stuck in local minima. Some examples are particle swarm optimization \cite{swarm}, evolution strategies \cite{nes}, or genetic algorithms \cite{genetic}.

    These gradient-free techniques have also been applied to the case of quantum circuit optimization. Some start from a fixed ansatz and only focus on optimizing the weights \cite{genetic2}. The chosen fixed initial architecture can be \textit{problem inspired}, meaning that the way the gates are placed in the circuit depends on the given problem, which is optimal for its solution. Some examples are Ansätze derived from the field of quantum chemistry \cite{qc1}, \cite{qc2} or combinatorial optimization, as the famous quantum approximate optimization algorithm (QAOA) \cite{qaoa1}, \cite{qaoa2}. In alternative, they could be \textit{problem agnostic}, meaning that the architecture is independent of the problem as the hardware efficient ansatz \cite{hard-eff}, often used because of its ease in physical hardware implementation.

    Other techniques instead attempt to find the optimal architecture for the task of interest, which means heuristically trying to place new gates to make the circuit perform better according to some carefully designed metrics. Inspired by the field of quantum chemistry and designed to gradually evolve a variational quantum eigensolver are the ADAPT-VQE algorithm \cite{ADAPT-VQE}, one of the first algorithms of this type to be proposed, or EVQE \cite{EVQE} where new gates are added smoothly, and an informed removal of redundant gates is performed. 
    
    Various algorithms have been proposed to try to optimise the architecture of a quantum circuit in order to match a target matrix \cite{tm1}, \cite{tm2}, \cite{tm3}, \cite{tm4}, \cite{tm5}. The techniques here proposed are not easily generalisable though to the task of optimising VQCs also because they make use of non-parametrised gates. 

    In this paper we propose a genetic algorithm which takes inspiration from the classical neuroevolution of augmenting topologies (NEAT) \cite{neat} proposed in 2002 by Stanley and Miikkulainen, which is categorized as a sexual algorithm since it employs the crossover technique. As it will be shown, here the architecture of the circuit as well as the weights are otpimized at the same time, and trough the speciation diversity is preserved, meaning that different areas of the search space are explored.

    An adaptation of the NEAT algorithm to the quantum case is proposed, having a generic multi-purpose and NISQ-friendly algorithm for solving a variety of tasks in the field of quantum machine learning: from the tasks closely inspired by quantum problems to the classical machine learning tasks. In the following sections we will explore in detail the quantum neat algorithm and show some of the settings where it has been tested and the results.

\section{NEAT for Variational Quantum Circuits}\label{sec:qneat}
    In this chapter, we present our algorithm, which consists of an adapted version of the Neuroevolution for Augmented Topologies for the case of quantum variational architectures. We will first consider the case of a free or a pre-defined structure in the architecture and then explain how the genome, crossover, mutation, and speciation process have been adapted.

    \subsection{The architecture}
    
    To adapt the architecture to the quantum case, we first distinguish between a constrained and a free architecture. We define a variational quantum circuit with a \textbf{free architecture} as a VQC where there is no constraint on the placements of rotation gates (ROTs) and controlled-not gates (CNOTs). In other words, ROTs gates are allowed to concatenate one after the other, and CNOTs can connect any two wires. In such a case, no regularity is encountered, and thus no concept of a layer is defined.
        
    A VQC with a \textbf{constrained architecture} is instead a circuit where, after the initial encoding layer, the ROTs and CNOTs follow a regular pattern. There are many architectures, namely architecture Ansätze, that can be used. Some common examples are the \textit{hardware efficient} architecture or the \textit{strongly entangling} architecture. We will take the last one as a reference and describe it in more detail since it's often used in literature when dealing with the type of problems we will test our algorithm.

\begin{figure}
    
    \begin{center}
    \begin{quantikz}[column sep=10pt, row sep={0.7cm,between origins}]
    \lstick{$\ket{0}$}& \gate[4,disable auto height, style={inner xsep=-4pt}]{\begin{array}{c} \text{U} \end{array}}  & \gate{R}\gategroup[4,steps=5,style={dashed,rounded corners, inner xsep=2pt}, background]{} & \ctrl{1} & \qw & \qw & \targ{} & \meter{} \vqwexplicit{1-1}{1-3}\\
    \lstick{$\ket{0}$} \qw & \qw & \gate{R} & \targ{} & \ctrl{1} & \qw & \qw & \meter{} \vqwexplicit{2-1}{2-3} \\
    \lstick{$\ket{0}$} \qw & \qw & \gate{R} & \qw & \targ{} & \ctrl{1} & \qw & \qw \vqwexplicit{3-1}{3-3}\\
    \lstick{$\ket{0}$} \qw & \qw & \gate{R} & \qw & \qw & \targ{} & \ctrl{-3} & \qw \vqwexplicit{4-1}{4-3}
    \end{quantikz}
    \end{center}
    
    \caption{Quantum circuit with the Strongly Entangling Layers architecture}
    \label{fig:constrained}
    \end{figure}
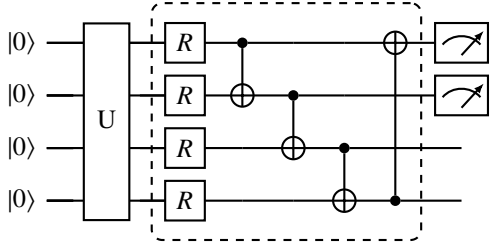

    After the usual encoding layer, the architecture consists of repeating layers of first a set of ROTs, one for each wire, and then a sequence of CNOTs, where each one of them goes from wire $i$ to wire $i+1$, $\mod(n)$, with $n$ the number of wires, as showed in Figure~\ref{fig:constrained}. The gates grouped into the dotted lines constitute one layer, which can be repeated for an arbitrary amount of time. In our case, the algorithm will determine how deep the circuit will be and, thus, how many layers it will be made of.
    
    \vspace{10px}

\begin{figure}
    \begin{center}

    \begin{quantikz}[column sep=8pt, row sep={0.7cm,between origins}]
    & \qw  & \gate{R}\gategroup[4,steps=5,style={dashed,rounded corners, inner xsep=0pt}, background]{{\sc 1}} & \ctrl{1} & \qw & \qw & \targ{} & \qw\gategroup[4,steps=4,style={dashed,rounded corners, inner xsep=-1.5pt}, background]{{\sc 2}} & \qw & \qw & \targ{} & \gate{R}\gategroup[4,steps=2,style={dashed,rounded corners, inner xsep=0pt}, background]{{\sc 3}} & \ctrl{1} & \meter{}\\
    \qw & \qw & \gate{R} & \targ{} & \ctrl{1} & \qw & \qw & \qw & \gate{R} & \ctrl{1} & \qw & \qw & \targ{} & \meter{}\vqwexplicit{2-1}{2-3}\\
    \qw & \qw & \gate{R} & \qw & \targ{} & \ctrl{1} & \qw & \qw & \gate{R} & \targ{} & \qw & \gate{R} & \qw & \qw \vqwexplicit{3-1}{3-3}\\
    \qw & \qw & \gate{R} & \qw & \qw & \targ{} & \ctrl{-3} & \qw & \qw & \qw & \ctrl{-3} & \qw & \qw & \qw \vqwexplicit{4-1}{4-3}
    \end{quantikz}
    
    \end{center}
    \caption{Example of a quantum constrained variational quantum circuit. Each layer is encircled with dotted lines. In each layer first the rotation gates and than the cnot gates are placed.}
    \label{fig:multiple-layers}
    \end{figure}
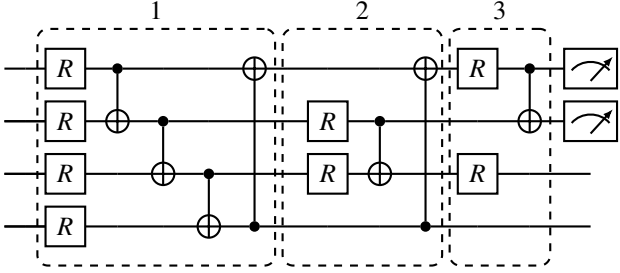

    To give e deeper insight into the resulting architecture once the algorithm made it evolve and added gates respecting the constraints, we look at one example in Figure~\ref{fig:multiple-layers}, where the encoding part has been left out for simplicity. The dotted lines identify the layers previously discussed. According to the constraints, these layers are not entirely filled but only partially. Layer number 1 is completely filled; layer 2 only has two ROTs and two CNOTs, while layer 3 is filled with two ROTs and one CNOT. In any case, we can see that the constraints are respected inside each layer: no more than one ROT for each wire is placed, and the CNOTs always connect wires $i$ and $ i+1, \mod(n)$. Inside each layer, we can thus identify a sublayer of ROTs and a sublayer of CNOTs.

    \vspace{10px}
	
    Clearly, the set of all possible architectures compatible with the constraints we posed is a subset of the ones that a free search could find. Nonetheless, the task would be computationally much more expensive. For example, only the number of CNOTs that could be placed in a layer grows with the number $n$ of wires as $\frac{n!}{2(n-2)!}$, while it grows linearly with $n$ in the constrained case. Also, no better performance on a practical level is guaranteed. For these reasons, from now on, we will only consider the case of a \textbf{constrained architecture} and leave the free case as a possible future research.

    \subsection{Genome}
    
    Each agent of the population is endowed with genetic information that encodes which gates are present at a given moment of the evolution and where they are placed. Through the genes of the agent, its architecture can be reconstructed. 
    
    If we consider a VQC with $n$ layers and $m$ wires, then each gate is uniquely identified by a tuple $(t, l, w)$ containing information about the type, layer, and wire of the gate. More specifically 
    \begin{equation*}
        t \in \{ \text{ROT}, \text{CNOT} \}, \ l \in \{1, \dots, n\}, and \ w \in \{1, \dots, m\}. 
    \end{equation*}
    This information is enough to identify any gate because in the constrained architecture, once a layer $l$ is specified, we also know to which sublayer the gate belongs, meaning that all the ROTs must be placed before the CNOTs. Parameter $w$ indicates which wire it belongs to, and if we are dealing with a ROT gate, then it can already be placed without any other ambiguity. Otherwise, if it's a CNOT gate, we know that it will go from $w \to w+1, \mod(m)$, which also determines uniquely where to place it. 
    
    The genes also contain the innovation numbers, which, analogously to the classical NEAT algorithm, keep track of the chronological time step in which the mutation happened and are later used to compare mutations that occurred at the same time for the crossover. Each innovation number uniquely identifies a tuple $(l, w)$. Since the architecture is constrained, ROTs and CNOTs are always placed in their respective sublayers, which means that a mutation producing a CNOT could not have produced a ROT in the same place. Thus the innovation numbers of CNOTs and ROTs should be compared separately. On this basis, we divide the genome lists depending on the type of gate. 
    
    Every ROT gate also contains information about the rotation angles. As an explicit example we show the genome of the architecture in Figure~\ref{fig:multiple-layers} in Table~\ref{tab:cnots-tab} and~\ref{tab:rots-tab}.

    \begin{table}[h!]
    \begin{center}
    \scalebox{0.8}{
        
        \def\arraystretch{1.5}
        \begin{tabular}{|c|c|c|c|c|c|c|c|c|}
            \hline
            Layer & 1 & 1 & 1 & 1 & 2 & 2 & 3 & 3  \\ 
            Wire &  1 & 2 & 3 & 4 & 2 & 3 & 1 & 3\\ 
            Angles & $\vec{\theta}_{11}$ & $\vec{\theta}_{12}$ & $\vec{\theta}_{13}$ & $\vec{\theta}_{14}$ & $\vec{\theta}_{22}$ & $\vec{\theta}_{23}$ & $\vec{\theta}_{31}$ & $\vec{\theta}_{33}$ \\
            Innov. n. & 1 & 2 & 3 & 4 & 6 & 7 & 9 & 10 \\
            \hline
        \end{tabular}}
    \end{center}
    \caption{Genome of ROTs for architecture in Figure~\ref{fig:multiple-layers}. The rotation angles $(\theta_x, \theta_y, \theta_z)$ are encoded in a vector $\vec{\theta}_{l,w}$ where $l,w$ stand for the layer and wire.}
    \label{tab:cnots-tab}
    \end{table}

    \begin{table}[h!]
    \begin{center}
    \scalebox{0.8}{
    \def\arraystretch{1.5}
        \begin{tabular}{|c|c|c|c|c|c|c|c|}
            \hline
            Layer & 1 & 1 & 1 & 1 & 2 & 2 & 3 \\ 
            Wire (from) &  1 & 2 & 3 & 4 & 2 & 4 & 1 \\ 
            (Wire to) & 2 & 3 & 4 & 1 & 3 & 1 & 2 \\
            Innov. n. & 1 & 2 & 3 & 4 & 6 & 8 & 9 \\
            \hline
        \end{tabular}
    }
    \end{center}
    \caption{Genome of CNOTs for architecture in Figure~\ref{fig:multiple-layers}. The genome of each CNOT contains the information of the wire on which the control depends on, here called \textit{Wire (from)}. The information about the wire it is acting on (\textit{Wire to}) is redundant because of the constrains on the architecture. Here is shown anyway for completeness.}
    \label{tab:rots-tab}
    \end{table}

    \subsection{Crossover}
    Crossover is the process through which the genome of two members of the population gets mixed during the reproduction of the fittest members to produce the offspring's genome. Given two genomes, we define \textbf{matching genes} the ones that, given a type $t \in \{\text{ROT}, \text{CNOT}\}$ have the same tuple $(l, w)$. In the case of ROTs, this means that two rotation gates, even though they may have different angles, are considered matching genes if placed at the same point of the circuit's architecture. In the case of CNOTs, instead, the above definition corresponds to saying that two CNOTs are matching if they connect identical qubits. We then define \textbf{disjoint genes} as those that differ for one of the values in the tuple $(l, w)$. 
    
    When two members reproduce, we want the fittest member's genome to be the basis for the future offspring, with some modifications deriving from the less fit one. The crossover method works by aligning the genomes with respect to the innovation numbers so that the chronology of mutations allows us to compare and see which are the matching or disjoint genes. Then we select the genes for the offspring with the following rules:
    \begin{enumerate}
        \item \textbf{Matching genes} are inherited randomly. 
        \item \textbf{Disjoint genes} are inherited from the fittest parent. If the two parents have the same fitness, disjoint genes will also be chosen randomly. 
    \end{enumerate}
    The crossover process is performed separately for CNOTs and ROTs since so are the genome lists. In Figure~\ref{fig:crossover}, we give an explicit example of crossover for two simple architectures. We assume that two architectures have the same fitness, and we highlight the layers with the dotted boxes.

    \begin{center}
    \begin{figure}[h]
    \begin{tikzpicture}
        \node[scale=0.8] {
            \tikzset{
                operator/.append style={fill=blue!20},
            }
            \begin{quantikz}[column sep=8pt, row sep={0.7cm,between origins}]
                & \qw & \gate{R}\gategroup[4,steps=3,style={dashed,rounded corners, inner xsep=-0.5pt}, background]{{\sc 1}} & \ctrl{1} & \qw & \qw\gategroup[4,steps=2,style={dashed,rounded corners, inner xsep=-0.5pt}, background]{{\sc 2}} & \targ{} & \meter{} \\
                & \qw & \qw & \targ{} & \qw & \gate{R} & \qw & \meter{} \\
                & \qw & \qw & \qw & \ctrl{1} & \qw & \qw & \qw \\
                & \qw & \qw & \qw & \targ{} & \qw & \ctrl{-3} & \qw
            \end{quantikz}
            $\; \; \; \; \; \; \varprod \; \; \; \; \; \;$
            \tikzset{
                operator/.append style={fill=white!20},
            }
            \begin{quantikz}[column sep=8pt, row sep={0.7cm,between origins}]
                & \qw & \gate{R}\gategroup[4,steps=5,style={dashed,rounded corners, inner xsep=-0.5pt}, background]{{\sc 1}} & \ctrl{1} & \qw & \qw & \targ{} & \meter{} \\
                & \qw & \qw & \targ{} & \ctrl{1} & \qw & \qw & \meter{} \\
                & \qw & \gate{R} & \qw & \targ{1} & \ctrl{1} & \qw & \qw \\
                & \qw & \qw & \qw & \qw{} & \targ{} & \ctrl{-3} & \qw
            \end{quantikz}
        };

        \node (A) at (0,-1.1){};
        \node (B) at (0,-2.1){};
        \draw[shorten >=0cm,shorten <= 0cm,->] (A)--node[auto, align=center]{}(B);

    
        \newcommand\squareX{3.5}
        \newcommand\squarey{-3.5}
        \newcommand\squareY{2.0}

        \newcommand\ssquareX{2.3}
        \newcommand\ssquarey{-6.2}
        \newcommand\ssquareY{2}

        \newcommand\sideX{0.6}
        \newcommand\sideY{0.4}         
        \newcommand\shiftonx{0.6025}
        \newcommand\shiftony{0.55}
 
        \newcommand{\shiftX}[1]{%
            #1 * \shiftonx - 3 * \shiftonx
        }

        \newcommand{\shiftY}[1]{%
            - #1 * \shiftony + 0.6
        }


        \newcommand\shifttopsquare{0.2}
        \newcommand\shifttopsquareX{-1.4}

        \node[text width=3cm, rotate=90] at (-2.0 + \shifttopsquareX,-2.5) {CNOTs};
        
        \draw (-1 * \squareX/2 + \shifttopsquareX,\squarey + \squareY/2 + \shifttopsquare) -- (\squareX/2 + \shifttopsquareX, \squarey + \squareY/2 + \shifttopsquare) -- (\squareX/2 + \shifttopsquareX, \squarey - \squareY/2 + \shifttopsquare) -- (-1* \squareX/2 + \shifttopsquareX, \squarey - \squareY/2 + \shifttopsquare) -- (-1 * \squareX/2 + \shifttopsquareX, \squarey + \squareY/2 + \shifttopsquare);
        

        \filldraw [fill=blue!20] (-\sideX/2 + \shiftX{1} + \shifttopsquareX, \squarey+\sideY/2 + \shiftY{0} + \shifttopsquare) rectangle (\sideX/2 + \shiftX{1} + \shifttopsquareX, \squarey-\sideY/2 + \shiftY{0} + \shifttopsquare) node[pos=.5, scale=0.4] {\begin{tabular}{c} Layer: 1 \\ $1 -> 2$ \end{tabular}};

        \filldraw [fill=blue!20] (-\sideX/2 + \shiftX{3} + \shifttopsquareX, \squarey+\sideY/2 + \shiftY{0} + \shifttopsquare) rectangle (\sideX/2 + \shiftX{3} + \shifttopsquareX, \squarey-\sideY/2 + \shiftY{0} + \shifttopsquare) node[pos=.5, scale=0.4] {\begin{tabular}{c} Layer: 1 \\ $3 -> 4$ \end{tabular}};
        
        \filldraw [fill=blue!20] (-\sideX/2 + \shiftX{5} + \shifttopsquareX, \squarey+\sideY/2 + \shiftY{0} + \shifttopsquare) rectangle (\sideX/2 + \shiftX{5} + \shifttopsquareX, \squarey-\sideY/2 + \shiftY{0} + \shifttopsquare) node[pos=.5, scale=0.4] {\begin{tabular}{c} Layer: 2 \\ $4 -> 1$ \end{tabular}};

        \filldraw [fill=white] (-\sideX/2 + \shiftX{1} + \shifttopsquareX, \squarey+\sideY/2 + \shiftY{1} + \shifttopsquare) rectangle (\sideX/2 + \shiftX{1} + \shifttopsquareX, \squarey-\sideY/2 + \shiftY{1} + \shifttopsquare) node[pos=.5, scale=0.4] {\begin{tabular}{c} Layer: 1 \\ $1 -> 2$ \end{tabular}};

        \filldraw [fill=white] (-\sideX/2 + \shiftX{2} + \shifttopsquareX, \squarey+\sideY/2 + \shiftY{1} + \shifttopsquare) rectangle (\sideX/2 + \shiftX{2} + \shifttopsquareX, \squarey-\sideY/2 + \shiftY{1} + \shifttopsquare) node[pos=.5, scale=0.4] {\begin{tabular}{c} Layer: 1 \\ $2 -> 3$ \end{tabular}};

        \filldraw [fill=white] (-\sideX/2 + \shiftX{3} + \shifttopsquareX, \squarey+\sideY/2 + \shiftY{1} + \shifttopsquare) rectangle (\sideX/2 + \shiftX{3} + \shifttopsquareX, \squarey-\sideY/2 + \shiftY{1} + \shifttopsquare) node[pos=.5, scale=0.4] {\begin{tabular}{c} Layer: 1 \\ $3 -> 4$ \end{tabular}};

        \filldraw [fill=white] (-\sideX/2 + \shiftX{4} + \shifttopsquareX, \squarey+\sideY/2 + \shiftY{1} + \shifttopsquare) rectangle (\sideX/2 + \shiftX{4} + \shifttopsquareX, \squarey-\sideY/2 + \shiftY{1} + \shifttopsquare) node[pos=.5, scale=0.4] {\begin{tabular}{c} Layer: 1 \\ $4 -> 1$ \end{tabular}};

        \node (A) at (-1.7 + \shifttopsquareX,-3.78 + \shifttopsquare){};
        \node (B) at (1.7 + \shifttopsquareX, -3.78 + \shifttopsquare){};
        \draw[] (A)--node[auto, align=center]{}(B);


        \filldraw [fill=white] (-\sideX/2 + \shiftX{1} + \shifttopsquareX, \squarey+\sideY/2 + \shiftY{2} - 0.1 + \shifttopsquare) rectangle (\sideX/2 + \shiftX{1} + \shifttopsquareX, \squarey-\sideY/2 + \shiftY{2} - 0.1 + \shifttopsquare) node[pos=.5, scale=0.4] {\begin{tabular}{c} Layer: 1 \\ $1 -> 2$ \end{tabular}};

        \filldraw [fill=white] (-\sideX/2 + \shiftX{2} + \shifttopsquareX, \squarey+\sideY/2 + \shiftY{2} - 0.1 + \shifttopsquare) rectangle (\sideX/2 + \shiftX{2} + \shifttopsquareX, \squarey-\sideY/2 + \shiftY{2} - 0.1 + \shifttopsquare) node[pos=.5, scale=0.4] {\begin{tabular}{c} Layer: 1 \\ $2 -> 3$ \end{tabular}};

        \filldraw [fill=white] (-\sideX/2 + \shiftX{3} + \shifttopsquareX, \squarey+\sideY/2 + \shiftY{2} - 0.1 + \shifttopsquare) rectangle (\sideX/2 + \shiftX{3} + \shifttopsquareX, \squarey-\sideY/2 + \shiftY{2} - 0.1 + \shifttopsquare) node[pos=.5, scale=0.4] {\begin{tabular}{c} Layer: 1 \\ $3 -> 4$ \end{tabular}};

        \filldraw [fill=blue!20] (-\sideX/2 + \shiftX{5} + \shifttopsquareX, \squarey+\sideY/2 + \shiftY{2} - 0.1 + \shifttopsquare) rectangle (\sideX/2 + \shiftX{5} + \shifttopsquareX, \squarey-\sideY/2 + \shiftY{2} - 0.1 + \shifttopsquare) node[pos=.5, scale=0.4] {\begin{tabular}{c} Layer: 2 \\ $4 -> 1$ \end{tabular}};

        \newcommand\shiftbottomsquare{2.9}
        \newcommand\shiftbottomsquareX{2.3}

        \node[text width=3cm, rotate=90] at (-1.4 + \shiftbottomsquareX,-5.2 +\shiftbottomsquare) {ROTs};

        \draw (-1 * \ssquareX/2 + \shiftbottomsquareX,\ssquarey + \ssquareY/2 + \shiftbottomsquare) -- (\ssquareX/2 + \shiftbottomsquareX, \ssquarey + \ssquareY/2 + \shiftbottomsquare) -- (\ssquareX/2 + \shiftbottomsquareX, \ssquarey - \ssquareY/2 + \shiftbottomsquare) -- (-1* \ssquareX/2 + \shiftbottomsquareX, \ssquarey - \ssquareY/2 + \shiftbottomsquare) -- (-1 * \ssquareX/2 + \shiftbottomsquareX, \ssquarey + \ssquareY/2 + \shiftbottomsquare);

        \newcommand{\shiftYbottom}[1]{%
            - #1 * \shiftony + 0.6 - 2.15
        }

        \filldraw [fill=blue!20] (-\sideX/2 + \shiftX{2} + \shiftbottomsquareX, \squarey+\sideY/2 + \shiftYbottom{1} + \shiftbottomsquare) rectangle (\sideX/2 + \shiftX{2} + \shiftbottomsquareX, \squarey-\sideY/2 + \shiftYbottom{1} + \shiftbottomsquare) node[pos=.5, scale=0.4] {\begin{tabular}{c} Layer: 1 \\ Wire: 1 \end{tabular}};

        \filldraw [fill=blue!20] (-\sideX/2 + \shiftX{4} + \shiftbottomsquareX, \squarey+\sideY/2 + \shiftYbottom{1} + \shiftbottomsquare) rectangle (\sideX/2 + \shiftX{4} + \shiftbottomsquareX, \squarey-\sideY/2 + \shiftYbottom{1} + \shiftbottomsquare) node[pos=.5, scale=0.4] {\begin{tabular}{c} Layer: 2 \\ Wire: 2 \end{tabular}};

        \filldraw [fill=white] (-\sideX/2 + \shiftX{2} + \shiftbottomsquareX, \squarey+\sideY/2 + \shiftYbottom{2} + \shiftbottomsquare) rectangle (\sideX/2 + \shiftX{2} + \shiftbottomsquareX, \squarey-\sideY/2 + \shiftYbottom{2} + \shiftbottomsquare) node[pos=.5, scale=0.4] {\begin{tabular}{c} Layer: 1 \\ Wire: 1 \end{tabular}};
        
        \filldraw [fill=white] (-\sideX/2 + \shiftX{3} + \shiftbottomsquareX, \squarey+\sideY/2 + \shiftYbottom{2} + \shiftbottomsquare) rectangle (\sideX/2 + \shiftX{3} + \shiftbottomsquareX, \squarey-\sideY/2 + \shiftYbottom{2} + \shiftbottomsquare) node[pos=.5, scale=0.4] {\begin{tabular}{c} Layer: 1 \\ Wire: 3 \end{tabular}};

        \node (A) at (-1.1 + \shiftbottomsquareX,-6.48 + \shiftbottomsquare){};
        \node (B) at (1.1 + \shiftbottomsquareX, -6.48 + \shiftbottomsquare){};
        \draw[] (A)--node[auto, align=center]{}(B);


        \filldraw [fill=white] (-\sideX/2 + \shiftX{2} + \shiftbottomsquareX, \squarey+\sideY/2 + \shiftYbottom{3} - 0.1 + \shiftbottomsquare) rectangle (\sideX/2 + \shiftX{2} + \shiftbottomsquareX, \squarey-\sideY/2 + \shiftYbottom{3} - 0.1 + \shiftbottomsquare) node[pos=.5, scale=0.4] {\begin{tabular}{c} Layer: 1 \\ Wire: 1 \end{tabular}};

        \filldraw [fill=blue!20] (-\sideX/2 + \shiftX{4} + \shiftbottomsquareX, \squarey+\sideY/2 + \shiftYbottom{3} - 0.1 + \shiftbottomsquare) rectangle (\sideX/2 + \shiftX{4} + \shiftbottomsquareX, \squarey-\sideY/2 + \shiftYbottom{3} - 0.1 + \shiftbottomsquare) node[pos=.5, scale=0.4] {\begin{tabular}{c} Layer: 2 \\ Wire: 2 \end{tabular}};

        \node (A) at (0,-4.5){};
        \node (B) at (0,-5.5){};
        \draw[shorten >=0cm,shorten <= 0cm,->] (A)--node[auto, align=center]{}(B);

        \node[scale=.7][below=-0.3cm of B] {
            \tikzset{
                operator/.append style={fill=cyan!20},
            }
            \begin{quantikz}[column sep=8pt, row sep={0.7cm,between origins}]
                & \qw & \gate{R}\gategroup[4,steps=4,style={dashed,rounded corners, inner xsep=-0.5pt}, background]{{\sc 1}} & \ctrl{1} & \qw & \qw & \qw\gategroup[4,steps=2,style={dashed,rounded corners, inner xsep=-0.5pt}, background]{{\sc 2}}  & \targ{} & \meter{} \\
                & \qw & \qw & \targ{} & \ctrl{1} & \qw & \gate{R} &  \qw & \meter{} \\
                & \qw & \qw & \qw & \targ{} & \ctrl{1} & \qw & \qw & \qw \\
                & \qw & \qw & \qw & \qw & \targ{} & \qw & \ctrl{-3} & \qw
            \end{quantikz}
        };
        
    \end{tikzpicture}
    \caption{Crossover process for two simple VQCs where the encoding layer has been omitted. The crossover process takes place separetly for CNOTs and ROTs. In each box the genes are aligned with respect to the innovation number. The fist line lists the genome of the left VQC; the second line the genome of the right VQC; the third line the genome of the offspring, where the gates have been selected according to the rules explained. Here the two VQC are supposed to have the same fitness, thus the choice of matching or disjoint genes is made randomly.}
    \label{fig:crossover}
    \end{figure}
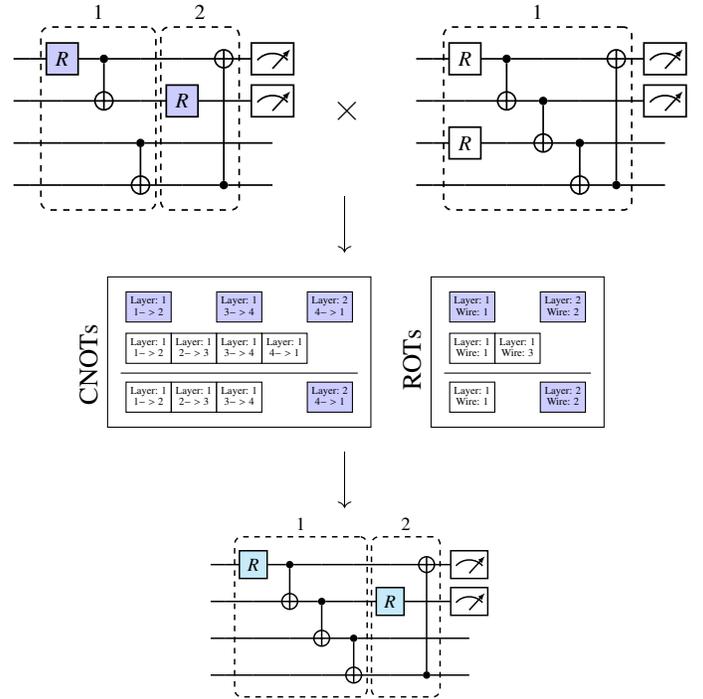
    \end{center}

    \subsection{Mutation}
    After the crossover has been performed, a mutation can take place. In the mutation process, there are three possible features that we can modify: The weights of the rotations can be adjusted (adding, for example, a small value sampled from a normal distribution); A new gate (ROT or CNOT) can be added respecting the rules of the constrained architecture. All the recent changes are then encoded into the genome.

    \subsection{Speciation}
    Furthermore, just like in the original idea of the NEAT algorithm, species are introduced to diversify the different evolutions and preserve the diversity for some time.  This is because a new architecture may perform well only once the weights have been appropriately adapted, which may take some evolution steps. 
    
    We thus divide the population into species, and, in order to understand if a generic agent is a member of a given species, we use the following metric to compare the agent with the best-performing member of the species.
    
    \begin{equation}
        \delta = c_1\frac{E}{N} + c_2\frac{D}{N} + c_3 \overline{W}, 
    \end{equation}
    
    where $c_1, c_2, c_3$ are linear coefficients that can be chosen arbitrarily, $E$ is the number of genes in excess, so the ones whose invention number is not reached by the other agent, $D$ is the number of disjoint genes, as defined before. $\overline{W}$ is the average distance between the rotation angles of matching genes. $N$ is the number of genes of the architecture containing the longest genome.
    
    The reproduction process is thus performed inside a single species. If a species is performing poorly, we want to penalize it by reducing its members, and the opposite if, on average, it's producing fit members. To do so, we consider species $j$ containing in a given moment $N_j$ members. The number of members of that species after the reproduction process will be 
    \begin{equation}
        N'_j = \sum\limits_{i = 0}^{N_j} \frac{f_{ij}}{\bar{f}}, 
    \end{equation}
    where $\bar{f}$ is the average fitness of the whole population.

\section{The algorithm}\label{sec:algorithm}
    We present the complete algorithm of QNEAT. The algorithm is explicitly described in Algorithm~\ref{alg:QNEAT}. The primary hyperparameters of the algorithm are the followings: the number of generations we want to run it for $N_g$, the population size $N$, the weights mutation coefficient, e.g., the standard deviation of the normal distribution $\mathcal{N}(0, \sigma)$ from which we will sample to change the rotation weights. Moreover, we define the probability that the weights of a gate will be changed as $p_w$, that a ROT gate will be added as $p_{ROT}$, and that a CNOT gate will be added as $p_{CNOT}$, the compatibility threshold $\delta_0$ to determine if a member is or is not part of a species. 

    One last crucial hyperparameter given in input is the number of initial total layers $i_L$. While we could start from an empty circuit and let QNEAT evolve it, as we will do in the experiments, another option is to start from a circuit with an initial number of layers filled in the sense of Figure~\ref{fig:constrained}. Initial layers may lead to a faster convergence from observation of experimental results, even though it's not guaranteed.


    \SetKwInput{KwInput}{Input}
    \RestyleAlgo{ruled}

    \begin{algorithm}[h]
    \caption{QNEAT Algorithm}\label{alg:QNEAT}
    
        \KwInput{
            $N_g, N, \sigma, p_w, p_{ROT}, p_{CNOT}, \delta_0, i_L$
        }
    
        \Begin{
            Create random population of size $N$ \\
    
            \For{every generation}{
                \tcc{Evaluate fitness}
                \For{every agent of the population}{
                    Evaluate the fitness of the agent 
                }
            }
        
            \tcc{Speciation}
            \For{every agent of the population}{ 
                \For{every species}{
                    Select best agent of species
                
                    $\delta \gets d(agent, \text{best agent})$
                
                    \If{$\delta < \delta_0$}{
                        Add agent to the current species 
                    }
                    
                    \Else{
                        Add a new species with current agent
                    }
                }
            }
            
            \tcc{Crossover}
            
            \For{every species}{
                Select the best performing agents and kill the others\\
                Calculate the number of necessary spawns $N'$ \\
                
                \For{$N'$ times}{
                    Select random agent1 and agent2 between the most performing ones \\
                    Child $\gets$ $\text{ } crossover(agent1, agent2)$ \\
                    \tcc{Mutation}
                    Child $\gets mutate(child)$ \\
                    Add child to the species
                }
            }
        }
    \end{algorithm}

\section{Experiments settings}\label{sec:experiments}
    To explore the range of tasks the algorithm can be applied to, we test it on a variety of benchmark problems. Namely, the algorithm has been tried on reinforcement learning and optimization tasks. In each task of benchmarks, the algorithm remains the same apart from the function we use to evaluate the fitness of population members.

    \subsection{Reinforcement Learning}
        The field of quantum reinforcement learning is rapidly evolving \cite{qrl}. Different techniques are being employed, each one with a different degree of quantum-classical hybridization. As a reference we take the case of deep reinforcement learning, where the well known Q-value method is solved with a function approximator: a classical neural network is used to learn the $Q$ function, so the expected future total reward, by an update rule that comes from the Bellman equation. This leads the function $Q$ to its desired value:
        
        \begin{equation}
            Q(s, a) = \mathbb{E}\left[ r + \gamma \max\limits_{a \in \mathcal{A}} Q(s', a) \right]
        \end{equation}

        where $a \in \mathcal{A}$ is an action in the action space, $s, s' \in \mathcal{S}$ are the current and next observation in the observation space, $r$ the reward of the action taken and $\gamma$ the discount factor. 
        
        Different attempts have been made to reproduce the same algorithm using a VQC as a function approximator for the $Q$ function \cite{qrl2}, \cite{qrl3}, \cite{qrl4}. Starting from this framework, we also want the QNEAT algorithm to learn the $Q$ value function. Since no gradient-update is employed here, in no point of the algorithm the constraint of the Bellman equation is applied, and thus the policy found may not respect it. 
        We can conclude thus that in this context the QNEAT algorithm behaves like an informed random search for a good (supposedly optimal) policy. 
        
        We test the algorithm on the Cart Pole and Frozen Lake benchmarks.
        In the first case we encode the observation space information in the qubits by a simple angle rotation with respect to the $x$ axis: for the $i$-th observation, whose value we  call $\theta$, a rotation $R_x(\theta)$ is applied to the gate $i$. We measure as many qubits as the dimension of the action space. In the Cart Pole problem the dimensions of the observation and action space are, respectively, four and two. 
        
        In the Frozen Lake benchmark we have an 8x8 grid for a total of 64 squares. We identify each square with a number from 1 to 64 and convert the number into a bit string of length $6 = \log_2(64)$. We thus use 6 qubits to encode the information applying to the $i$-th qubit a rotation of angle 
        \begin{equation}
            \theta_i = \pi x_i, \ \ \text{with} \;\; x_i \in \{0,1\}.
        \end{equation}

    \subsection{Optimization}
    As an example of optimization, we consider combinatorial optimization problems. In particular, we study the MaxCut problem, which is particularly important since many quadratic unconstrained binary optimization problems can be mapped into the MaxCut problem. Being able to solve it means thus the ability to solve a broader class of combinatorial optimization problems. 
    
    One traditional approach to solving combinatorial optimization problems is the Quantum Approximate Optimization Algorithm (QAOA)~\cite{qaoa1}. The QAOA can be seen as a particular case of Variational Quantum Eigensolvers (VQEs), a class of VQCs used to approximate the Schrödinger equation to evolve an initial state into a final one under an evolving Hamiltonian. 
    
    More specifically, given an initial Hamiltonian $H_i$ with a known ground state $\psi_{i,0}$, and a final Hamiltonian $H_f$ encoding our optimization problem with ground state $\psi_{f, 0}$ encoding the optimal solution, then we can start from the state $\psi_{i,0}$ and make it evolve with an appropriate unitary operation into the final one 
    \begin{equation}
        \ket{\psi_{f,0}} = U \ket{ \psi_{i, 0} }, 
    \end{equation}
    where the action of the unitary operator $U$ is performed by the VQC. We can thus map this problem into a VQC by encoding the initial ground state $\ket{\psi_{i, 0}}$ into the qubits and find an appropriate architecture to simulate a unitary time evolution $U$ to the final state encoding the optimal solution.
    
    To give a concrete example, one possible choice of $U$ is the one given by the Schrödinger equation. Namely, we can choose an evolving Hamiltonian $H(t) = tH_f + (1-t)H_i$, with $H_f$ the one encoding our problem. Then we can exponentiate it to obtain the Schrödinger equation
    \begin{equation}
        \ket{\psi_{f,0}} = \exp{\frac{- i H(t) t}{\hbar}} \ket{ \psi_{i, 0} }, 
    \end{equation}
    where the exponential can be decomposed analytically into physical gates, and a precise architecture can be found. This is only one possible choice of architecture, and many have been proposed \cite{qveigen}. Here we use the QNEAT algorithm to evolve the circuit and evolutionarily find the optimal architecture. 
    
    In this class of problems, we aim to find the state that encodes the lowest energy of the problem Hamiltonian $H_f$. In other words, we want to minimize the following expectation value
    \begin{equation}
        \expval{H_f}{\psi_f}. 
    \end{equation}
    The MaxCut problem we took into account consists of dividing the nodes on a graph into two separate sets so that the imaginary line drawn to encircle one, or the other set goes through the maximum number of edges in the graph. We can label the nodes $0$ or $1$ depending on the set they belong to, and these values can be encoded into the qubits. With an appropriate Hamiltonian, we can encode the information of the edge between the two being cut or not.

    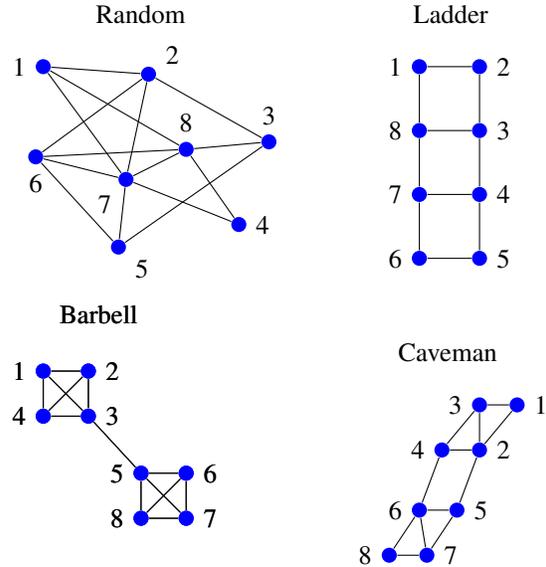
\begin{figure}
        \centering
    \begin{tikzpicture}
    \node[text width=3cm] at (2.2, 0.7) {Random};
    
        \node[shape=circle, fill=blue, minimum size = 0.2cm, inner sep=0pt, label=left:{$1$}] (0) at (-0.0, -0) {};
        \node[shape=circle, fill=blue, minimum size = 0.2cm, inner sep=0pt, label=5:{$2$}] (1) at (1.4, -0.1) {};
        \node[shape=circle, fill=blue, minimum size = 0.2cm, inner sep=0pt, label=above:{$3$}] (2) at (3.0, -1.0) {};
        \node[shape=circle, fill=blue, minimum size = 0.2cm, inner sep=0pt, label=right:{$4$}] (3) at (2.6, -2.1) {};
        \node[shape=circle, fill=blue, minimum size = 0.2cm, inner sep=0pt, label=-20:{$5$}] (4) at (1.0, -2.4) {};
        \node[shape=circle, fill=blue, minimum size = 0.2cm, inner sep=0pt, label=below:{$6$}] (5) at (-0.1, -1.2) {};
        \node[shape=circle, fill=blue, minimum size = 0.2cm, inner sep=0pt, label=230:{$7$}] (6) at (1.1, -1.5) {};
        \node[shape=circle, fill=blue, minimum size = 0.2cm, inner sep=0pt, label=above:{$8$}] (7) at (1.9, -1.1) {};

        \path [-] (0) edge node[left] {} (1);
        \path [-] (0) edge node[left] {} (6);
        \path [-] (0) edge node[left] {} (7);
        
        \path [-] (1) edge node[left] {} (2);
        \path [-] (1) edge node[left] {} (6);
        
        \path [-] (2) edge node[left] {} (7);
        \path [-] (2) edge node[left] {} (4);

        \path [-] (3) edge node[left] {} (7);
        \path [-] (3) edge node[left] {} (6);

        \path [-] (4) edge node[left] {} (6);
        \path [-] (4) edge node[left] {} (5);

        \path [-] (5) edge node[left] {} (1);
        \path [-] (5) edge node[left] {} (6);
        \path [-] (5) edge node[left] {} (7);

        \path [-] (6) edge node[left] {} (7);

        \newcommand\shiftX{5}
        \newcommand\shiftY{0}
        
        \node[text width=3cm] at (\shiftX + 1.42, \shiftY + 0.7) {Ladder};

        \node[shape=circle, fill=blue, minimum size = 0.2cm, inner sep=0pt, label=left:{$1$}] (0) at (\shiftX + 0, \shiftY + 0) {};
        \node[shape=circle, fill=blue, minimum size = 0.2cm, inner sep=0pt, label=right:{$2$}] (1) at (\shiftX + 0.8, \shiftY + 0) {};

        \node[shape=circle, fill=blue, minimum size = 0.2cm, inner sep=0pt, label=left:{$8$}] (7) at (\shiftX + 0, \shiftY - 0.85) {};
        \node[shape=circle, fill=blue, minimum size = 0.2cm, inner sep=0pt, label=right:{$3$}] (2) at (\shiftX + 0.8, \shiftY - 0.85) {};

        \node[shape=circle, fill=blue, minimum size = 0.2cm, inner sep=0pt, label=left:{$7$}] (6) at (\shiftX + 0, \shiftY - 1.7) {};
        \node[shape=circle, fill=blue, minimum size = 0.2cm, inner sep=0pt, label=right:{$4$}] (3) at (\shiftX + 0.8, \shiftY - 1.7) {};

        \node[shape=circle, fill=blue, minimum size = 0.2cm, inner sep=0pt, label=left:{$6$}] (5) at (\shiftX + 0, \shiftY - 2.55) {};
        \node[shape=circle, fill=blue, minimum size = 0.2cm, inner sep=0pt, label=right:{$5$}] (4) at (\shiftX + 0.8, \shiftY - 2.55) {};

        \path [-] (0) edge node[left] {} (1);
        \path [-] (7) edge node[left] {} (2);
        \path [-] (6) edge node[left] {} (3);        
        \path [-] (5) edge node[left] {} (4);

        \path [-] (0) edge node[left] {} (7);
        \path [-] (7) edge node[left] {} (6);
        \path [-] (6) edge node[left] {} (5);

        \path [-] (1) edge node[left] {} (2);
        \path [-] (2) edge node[left] {} (3);
        \path [-] (3) edge node[left] {} (4);

        \newcommand\shiftXb{0}
        \newcommand\shiftYb{-4}

        \node[text width=3cm] at (\shiftXb + 1.72, \shiftYb + 0.7) {Barbell};
            
        \node[shape=circle, fill=blue, minimum size = 0.2cm, inner sep=0pt, label=left:{$1$}] (0) at (\shiftXb + 0, \shiftYb - 0.05) {};
        \node[shape=circle, fill=blue, minimum size = 0.2cm, inner sep=0pt, label=right:{$2$}] (1) at (\shiftXb + 0.6, \shiftYb - 0.05) {};
        \node[shape=circle, fill=blue, minimum size = 0.2cm, inner sep=0pt, label=left:{$4$}] (3) at (\shiftXb + 0, \shiftYb - 0.65) {};
        \node[shape=circle, fill=blue, minimum size = 0.2cm, inner sep=0pt, label=right:{$3$}] (2) at (\shiftXb + 0.6, \shiftYb - 0.65) {};
        \node[shape=circle, fill=blue, minimum size = 0.2cm, inner sep=0pt, label=left:{$5$}] (4) at (\shiftXb + 1.3, \shiftYb - 0.65/0.6*1.3) {};
        \node[shape=circle, fill=blue, minimum size = 0.2cm, inner sep=0pt, label=right:{$7$}] (6) at (\shiftXb + 1.9, \shiftYb - 0.65/0.6*1.9 + 0.05) {};
        \node[shape=circle, fill=blue, minimum size = 0.2cm, inner sep=0pt, label=right:{$6$}] (5) at (\shiftXb + 1.9, \shiftYb - 0.65/0.6*1.3) {};
        \node[shape=circle, fill=blue, minimum size = 0.2cm, inner sep=0pt, label=left:{$8$}] (7) at (\shiftXb + 1.3, \shiftYb - 0.65/0.6*1.9 + 0.05) {};

        \path [-] (0) edge node[left] {} (1);
        \path [-] (0) edge node[left] {} (3);
        \path [-] (0) edge node[left] {} (2);        
        \path [-] (1) edge node[left] {} (3);
        \path [-] (1) edge node[left] {} (2);
        \path [-] (3) edge node[left] {} (2);
        
        \path [-] (2) edge node[left] {} (4);
        
        \path [-] (4) edge node[left] {} (5);
        \path [-] (4) edge node[left] {} (7);
        \path [-] (4) edge node[left] {} (6);        
        \path [-] (5) edge node[left] {} (7);
        \path [-] (5) edge node[left] {} (6);
        \path [-] (7) edge node[left] {} (6);

        \newcommand\barbellShiftXb{0}
        \newcommand\barbellShiftYb{-4}

        \node[text width=3cm] at (\shiftXb + 1.72, \shiftYb + 0.7) {Barbell};
            
        \node[shape=circle, fill=blue, minimum size = 0.2cm, inner sep=0pt, label=left:{$1$}] (0) at (\barbellShiftXb + 0, \barbellShiftYb - 0.05) {};
        \node[shape=circle, fill=blue, minimum size = 0.2cm, inner sep=0pt, label=right:{$2$}] (1) at (\barbellShiftXb + 0.6, \barbellShiftYb - 0.05) {};
        \node[shape=circle, fill=blue, minimum size = 0.2cm, inner sep=0pt, label=left:{$4$}] (3) at (\barbellShiftXb + 0, \barbellShiftYb - 0.65) {};
        \node[shape=circle, fill=blue, minimum size = 0.2cm, inner sep=0pt, label=right:{$3$}] (2) at (\barbellShiftXb + 0.6, \barbellShiftYb - 0.65) {};
        \node[shape=circle, fill=blue, minimum size = 0.2cm, inner sep=0pt, label=left:{$5$}] (4) at (\barbellShiftXb + 1.3, \barbellShiftYb - 0.65/0.6*1.3) {};
        \node[shape=circle, fill=blue, minimum size = 0.2cm, inner sep=0pt, label=right:{$7$}] (6) at (\barbellShiftXb + 1.9, \barbellShiftYb - 0.65/0.6*1.9 + 0.05) {};
        \node[shape=circle, fill=blue, minimum size = 0.2cm, inner sep=0pt, label=right:{$6$}] (5) at (\barbellShiftXb + 1.9, \barbellShiftYb - 0.65/0.6*1.3) {};
        \node[shape=circle, fill=blue, minimum size = 0.2cm, inner sep=0pt, label=left:{$8$}] (7) at (\barbellShiftXb + 1.3, \barbellShiftYb - 0.65/0.6*1.9 + 0.05) {};

        \path [-] (0) edge node[left] {} (1);
        \path [-] (0) edge node[left] {} (3);
        \path [-] (0) edge node[left] {} (2);        
        \path [-] (1) edge node[left] {} (3);
        \path [-] (1) edge node[left] {} (2);
        \path [-] (3) edge node[left] {} (2);
        
        \path [-] (2) edge node[left] {} (4);
        
        \path [-] (4) edge node[left] {} (5);
        \path [-] (4) edge node[left] {} (7);
        \path [-] (4) edge node[left] {} (6);        
        \path [-] (5) edge node[left] {} (7);
        \path [-] (5) edge node[left] {} (6);
        \path [-] (7) edge node[left] {} (6);

        \newcommand\shiftXc{4.5}
        \newcommand\shiftYc{-4.5}

        \node[text width=3cm] at (\shiftXc + 1.72, \shiftYc + 0.7) {Caveman};
            
        \node[shape=circle, fill=blue, minimum size = 0.2cm, inner sep=0pt, label=right:{$1$}] (0) at (\shiftXc + 1.8, \shiftYc - 0) {};
        \node[shape=circle, fill=blue, minimum size = 0.2cm, inner sep=0pt, label=left:{$3$}] (2) at (\shiftXc + 1.3, \shiftYc - 0) {};

        \node[shape=circle, fill=blue, minimum size = 0.2cm, inner sep=0pt, label=right:{$2$}] (1) at (\shiftXc + 1.3, \shiftYc - 0.6) {};
        \node[shape=circle, fill=blue, minimum size = 0.2cm, inner sep=0pt, label=left:{$4$}] (3) at (\shiftXc + 0.8, \shiftYc - 0.6) {};

        \node[shape=circle, fill=blue, minimum size = 0.2cm, inner sep=0pt, label=right:{$5$}] (4) at (\shiftXc + 1.0, \shiftYc - 1.4) {};
        \node[shape=circle, fill=blue, minimum size = 0.2cm, inner sep=0pt, label=left:{$6$}] (5) at (\shiftXc + 0.5, \shiftYc - 1.4) {};

        \node[shape=circle, fill=blue, minimum size = 0.2cm, inner sep=0pt, label=right:{$7$}] (6) at (\shiftXc + 0.6, \shiftYc - 2) {};
        \node[shape=circle, fill=blue, minimum size = 0.2cm, inner sep=0pt, label=left:{$8$}] (7) at (\shiftXc + 0.1, \shiftYc - 2) {};

        \path [-] (0) edge node[left] {} (2);
        \path [-] (0) edge node[left] {} (1);
        \path [-] (1) edge node[left] {} (2);

        \path [-] (2) edge node[left] {} (3);
        \path [-] (1) edge node[left] {} (3);

        \path [-] (1) edge node[left] {} (4);
        \path [-] (3) edge node[left] {} (5);

        \path [-] (4) edge node[left] {} (5);
        \path [-] (5) edge node[left] {} (6);
        \path [-] (4) edge node[left] {} (6);

        \path [-] (5) edge node[left] {} (7);
        \path [-] (7) edge node[left] {} (6);

    \end{tikzpicture}
        \caption{Benchmark graphs used for testing the performance of QNEAT on combinatorial optimization tasks.}
        \label{fig:graphs}
    \end{figure}

    The benchmark graphs used are shown in Figure~\ref{fig:graphs} and have been taken as example benchmarks from \cite{graphspaper}. For fitness we use the average value of the Hamiltonian of MaxCut, namely:
    \begin{equation}
        H = \sum\limits_{ (i,j) \in G} \frac{1}{2}\left( 1 - Z_i\otimes Z_j \right), 
    \end{equation}
    considering $Z_k$ the spin measured on the $k-$th qubit and $G$ being the set of the edges of the graph. Considering the output of the measure on the qubit to be in $\{-1, 1\}$, then if two nodes $(i, j)$ belong to the same set, so they have values $(1, 1)$ or $(-1, -1)$, then the edge contributes with a factor of 0 to the Hamiltonian, which corresponds to the fact that no cut goes through the edge. If instead they have different values, $(-1, 1)$ or $(1, -1)$, then the edge contributes with a factor 1. We thus aim at maximizing the Hamiltonian. 
    
    At the end of the evolution process, we select the best member and sample 100 times to see the distribution on the outcome results, so how many times the circuit gives a measured bit-string corresponding to one of the optimal solutions of the MaxCut problem for the given graph.

\section{Results}\label{sec:results}

   We present in Table~\ref{tab:hyper_init} the hyperparameters used to solve the different problems. With \textit{Reinforcement learning} we include both CartPole and FrozenLake8x8. With MaxCut we refer to all four graphs. The parameters have been chosen by doing a grid search.

    \begin{table}[h!]
    \centering
    \begin{tabular}{lrr}
        \toprule
        Param.  & Reinforcement learning & MaxCut    \\
        \midrule
        $N$       & 150             & 200        \\
        $\sigma$  & 0.01            & 0.01       \\
        $p_w$     & 0.3             & 0.6        \\
        $p_{ROT}$ & 0.5             & 0.3        \\
        $p_{CNOT}$& 0.5             & 0.3        \\
        $\delta_0$& 1.0             & 0.45       \\
        $i_L$     & 0/1/2           & 2          \\
        \bottomrule
    \end{tabular}
    \caption{Hyperparameters used to initialise the QNEAT algorithm to solve the various tasks: Reinforcement Learning (Cartpole and FrozenLake) and MaxCut (the four graphs)}
    \label{tab:hyper_init}
    \end{table}

    \subsection{Reinforcement Learning}
                
\begin{figure*}[h]
            \centering
            \includegraphics[width = 500 pt]{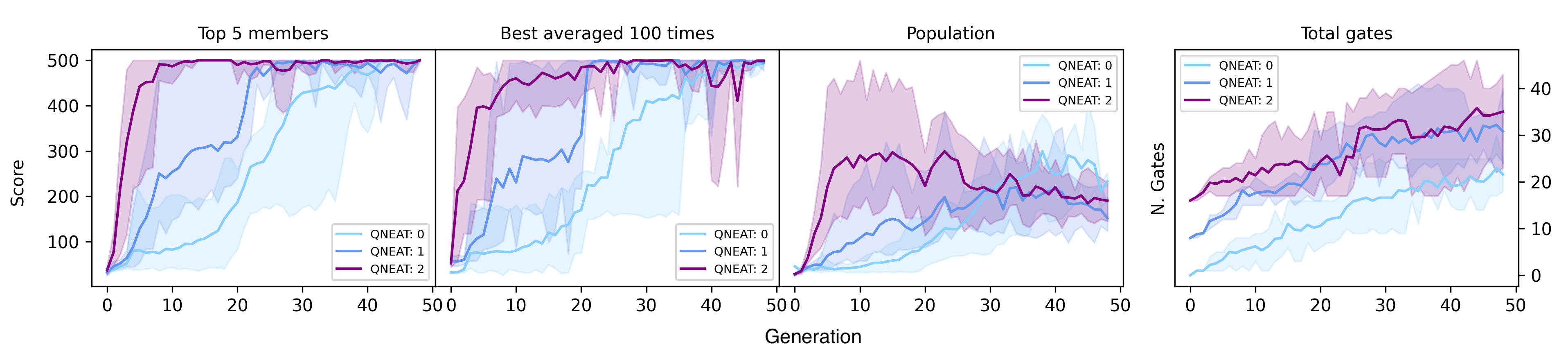}
            \caption{Results of the QNEAT algorithm for the Cart Pole environment. The performances of the top 5 agents, the best agent, and the whole population have been tracked. The best-performing agent has been tested 100 times. In the last column, the number of gates has been plotted to keep track of the complexity of the circuit.}
            \label{fig:cartpole-results}
        \end{figure*}

\begin{figure*}[h]
            \includegraphics[width = 500 pt]{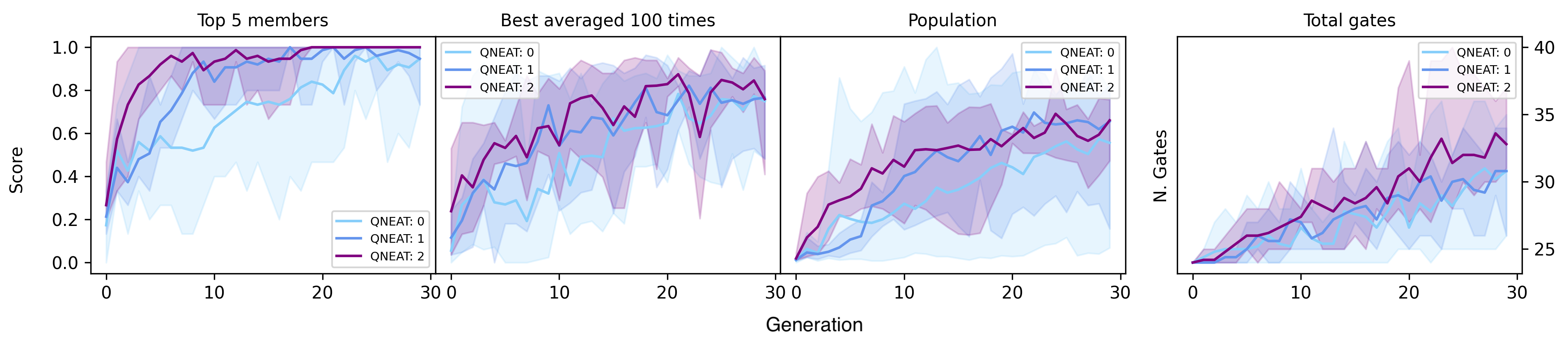}
            \caption{Results for the FrozenLake environment arranged in the same way described in the above plots.}
            \label{fig:frozenlake-results}
        \end{figure*}

        The results of Cart Pole and Frozen Lake are presented in Figure~\ref{fig:cartpole-results} and ~\ref{fig:frozenlake-results} respectively.
        We study the score of the VQE evolved by the QNEAT algorithm starting with $0, 1, 2$ initial layers.  
        We present separately the results of: the top 5 member, the top member averaged an extra 100 times and the whole population. The maximum score is 500.
        During the evolution, we also keep track of the number of ROTs and CNOTs that contribute to the architecture. 
        We can see how the top 5 agents rapidly reach the maximum score, converging faster when the number of initial layers increases. The population instead converges.

        \begin{figure*}[h]
            \centering
            \includegraphics[width = 500 pt]{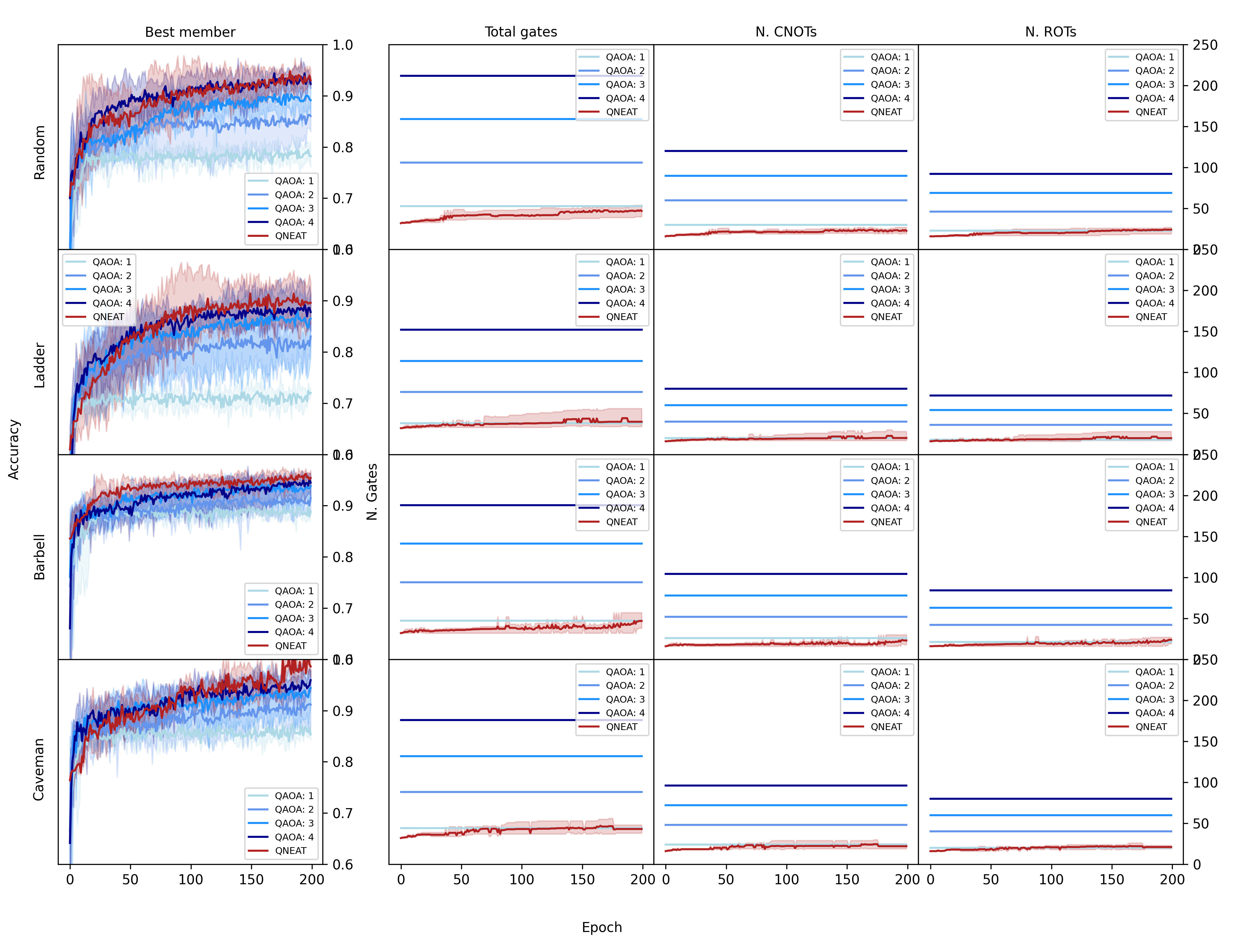}
            \caption{Results for the MaxCut problem. Each row describes one of the four graphs showed in Figure~\ref{fig:graphs}. The first column shows the accuracy while the other three show, respectively, the number of total gates, CNOTs and ROTs used for the architecture.}
            \label{maxcut-results}
        \end{figure*}
    
    \subsection{Combinatorial Optimization}
        
        We solve the MaxCut problem on the four different graphs that we have shown before. For each one of them, we evaluate the QAOA algorithm with 1, 2, 3, and 4 layers. Higher numbers of layers have also been tested, but they show the same behavior as the case with 4, and have thus been omitted. We compare it then with the QNEAT.
        In figure 6, the results can be seen. The four rows refer to the different graphs. In each row, we study the performance of the \textit{best member}: the first column shows the accuracy of measuring an optimal solution obtained by the QNEAT and QAOA algorithm.
        The last three rows show separately the number of total gates, CNOTs, and ROTs used in the architecture of each algorithm.
        Each epoch represents a generation of 200 members in the case of QNEAT, while only one gradient update in the case of the QAOA algorithm.
        From the graph, it can be easily seen how the QAOA improves with the number of layers used. This comes with a linear increase of gates used. Also, it can be seen how the QNEAT algorithm reaches good accuracies, comparable with the case of QAOA with four layers but with a much lower amount of gates. This makes the implementation more compatible with the current NISQ-era architectures.

    \subsection{Evolution}
\begin{figure*}[h]
            \centering
            \includegraphics[width = 450 pt]{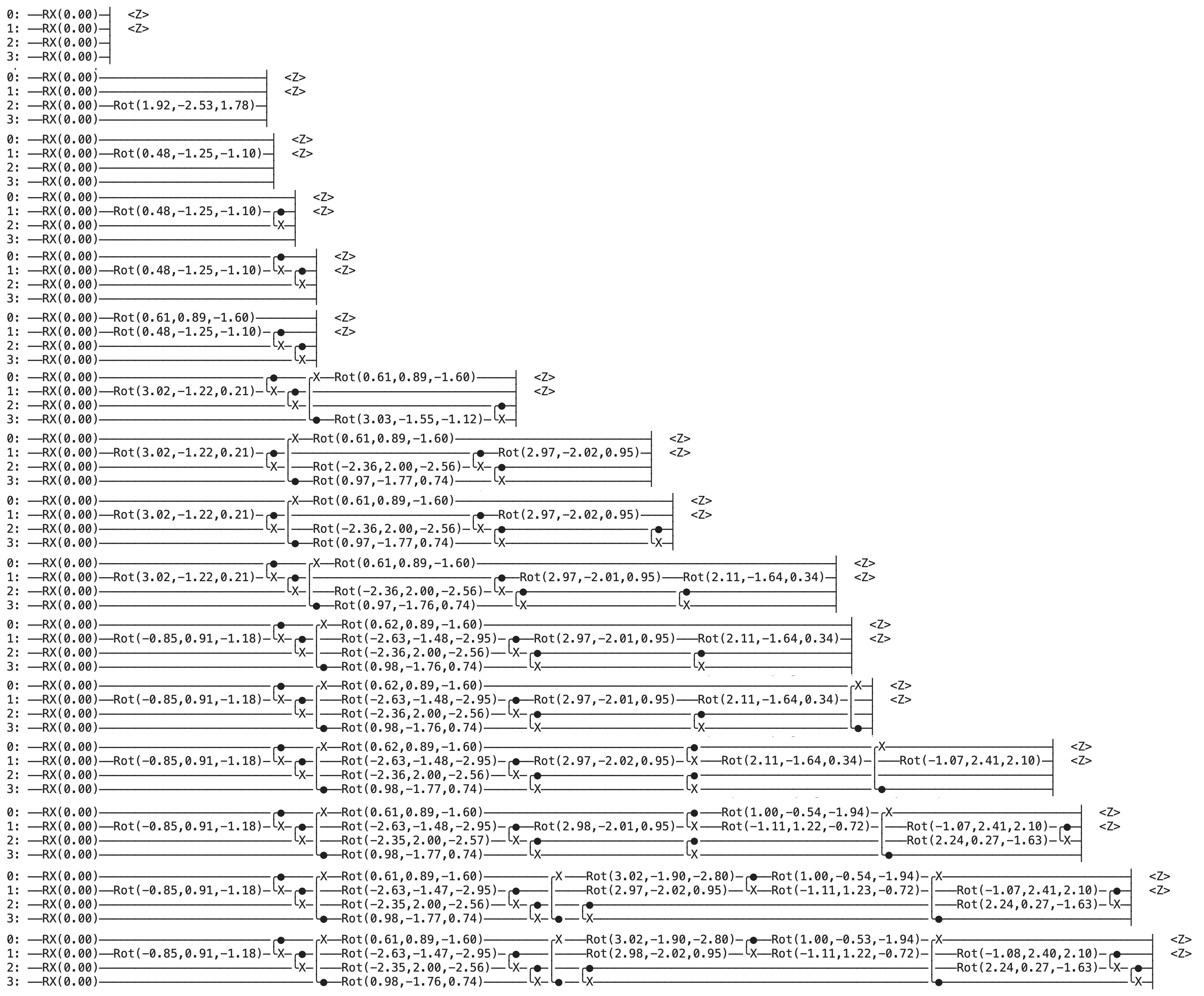}
            \caption{Evolution of an empty VQC with the QNEAT algorithm for the Cart Pole environment with parameters on Table~\ref{tab:hyper_init}. From top to bottom, each row is a different generation. Not all of the generations have been shown: of a total of 50 generations, only the ones where anew gate has been added. In the others, either no mutation happened or only the weights changed.}
            \label{fig:evolution}
        \end{figure*}

        We also show visually the evolution of one VQC starting from an empty configuration in the case of the Cart Pole benchmark. The parameters used for the next evolution are shown in Table~\ref{tab:hyper_init} for the Reinforcement Learning tasks. At the end of the evolution of the population, the best agent has been tracked backward and its architecture visualized generation by generation. Figure~\ref{fig:evolution} shows the evolution process. Only some of the 50 generations have been shown, namely the ones where new gates have been added. In the others, either nothing happened, or only the weights have been changed. 
        It can be seen that, starting from an empty architecture, apart from the encoding part, gates are slowly added until reaching slightly more than 20 gates, in accordance with the results shown in Figure~\ref{fig:cartpole-results}. In the case of the Cart Pole problem such a small architecture is capable of solving the benchmark and perform the top scores. 

\section{Multiple Objective Optimization}\label{sec:MOO}
    So far we have tested the QNEAT algorithm on the classical machine learning benchmark problems and we kept track of the number of gates that were added during the evolution. 

    In doing so we always relied on the fact that, because of how the QNEAT algorithm works, new gates are added only if the resulting performance is better, so only if it is necessary. We never explicitly requested, in any point of the algorithm, that the number of gates should be minimized. 
    
    Obviously, we want the number of gates to be as low as possible, provided that the agent is performing the top score of whatever the problem is considered. QNEAT algorithm, as it is, deals with this problem by only letting species with bigger architectures if they are performing better. Nonetheless, no penalty is added to deep, complex architectures. In order to enforce the algorithm to prioritize short architectures, we apply a technique known as \textit{Multi Objective Optimization} (MOO). 
    
    In particular, here we studied the NSGA-II algorithm \cite{nsga2}, an algorithm proposed to deal with optimization problems when having multiple scores or fitnesses to take account at the same time. 

    NSGA-II is a genetic algorithm based on the idea of finding the Pareto fronts in the space of the fitness functions, and then evolving the members of the Pareto fronts closer to the highest values of the scores.

    The algorithm works as follows: we start with a set of fitnesses $\{f_1, \dots, f_n\}$, which we suppose we want to minimize. This assumption can always be made true by simply inverting any eventual $f_i$ that should be maximized as $1/f_i$. Given a population of $N$ members, the algorithm starts with reproduction and mutation of the original population $P$ to generate the new set of members $Q$. The population will have thus $2N$ members, where half of them are the mutation of the initial ones. 
    
    After all of them have been let perform the tasks, we can plot each one as a point $(f_1, \dots, f_n) \in \mathbb{R}^n$ in the fitnesses space. 

    Here the set of points, representing the members, are divided in Pareto fronts, or \textit{nondomination fronts}. A member $A$ is said to dominate another member $B$ if all the fitnesses of $A$ are equal to the fitness of $B$, and at least one is strictly better. Once the different members have been compared, they can be divided into fronts, where each front has a rank. Each member of the front does not dominate and is not dominated by any of the other members; it dominates all the members of every other front with a lower rank, and is dominated by the members of higher-rank fronts. In different words, a domination front is the set of all those members that are roughly at the same distance with respect to the fitness axis, compared to the other members. 

    Once the population, now still counting $2N$ members, has been divided into Pareto fronts, we then order all the members according to the overall performance and select the first half of them to make the new population. The way the members are ordered in the end depends on two factors: the rank and the crowding distance.

    \begin{enumerate}
        \item \textbf{Rank}. Once the members are divided into Pareto fronts, each receives a rank, simply an integer number, corresponding to the Pareto front it belongs to. Once this has been done, all the members with a higher rank are considered to be better than those with a lower rank, and starting from these ones the mutation and crossover processes will take place.

        \item \textbf{Crowding distance}. Inside a nondomination front, where the rank is the same for every member, another metric is used to classify the members, the crowding distance. Given one element $i$, we define its crowding distance as the distance in the fitnesses space of the nearest members to him in the same front, averaged on all the fitnesses directions. The memebers who are in an area of the space that is more crowded are considered to be better.  
        
    \end{enumerate}

    More exactly, the condition to establish if a member $i$ is better than a member $j$ is 

    \begin{equation}\label{eq:condition}
        \begin{split}
            i < j \text{  if  } (i_{\text{rank}} < j_{\text{rank}}) \\
            \text{  or  } ( (i_{\text{rank}} = j_{\text{rank}}) \text{ and } (i_{\text{distance}} > j_{\text{distance}}))
        \end{split}
    \end{equation}

    According to the condition in equation \ref{eq:condition}, all the $2N$ members are ordered and then the first $N$ of them are selected to be part of the next generation. 

    The algorithm's main loop is shown in Algorithm~\ref{alg:NSGA2}, while the other algorithms mentioned inside can be consulted in the original work. 

    \begin{algorithm}[h]
        \caption{Multi Objective Optimization - NSGA-II}\label{alg:NSGA2}
  
        \Begin{
            $R_t = P_t \cup Q_t$ \\
            $\mathcal{F} = $ fast-non-dominated-sort($R_t$) \\
            $P_{t + 1} = \empty$ 
            
            \While{ $|P_{t+1}| + |\mathcal{F}_i| \leq N$ }{
                crowding-distance-sort($\mathcal{F}_i$) \\
                $P_{t + 1} = P_{t+1} \cup \mathcal{F}_i$ \\
                $i \gets i+1$
            }

            Sort ($\mathcal{F}_i$) \\
            $P_{t+1} \gets P_{t+1} \cup \mathcal{F}_i \left[ 1 : (N - |P_{t_1}|) \right]$ \\
            $Q_{t+1} = \text{make-new-pop}(P_{t+1})$
        }    
    \end{algorithm}

    In our case, we merged the QNEAT and MMO algorithms. More specifically, we kept the species structure of the first algorithms, and, inside each species, the MMO was applied, looking for the Pareto fronts separately for each species. This is done to keep the different species evolve independetly, so that the members inside each species don't need to compete with members from other species. 
    
    The mutation and crossover to generate new offsprings are instead drawn from the QNEAT algorithm.

    The hyperparameters used to run the algorithm are the same as in Table~\ref{tab:hyper_init}. 

\begin{figure}[t]
        \includegraphics[width = 250 pt]{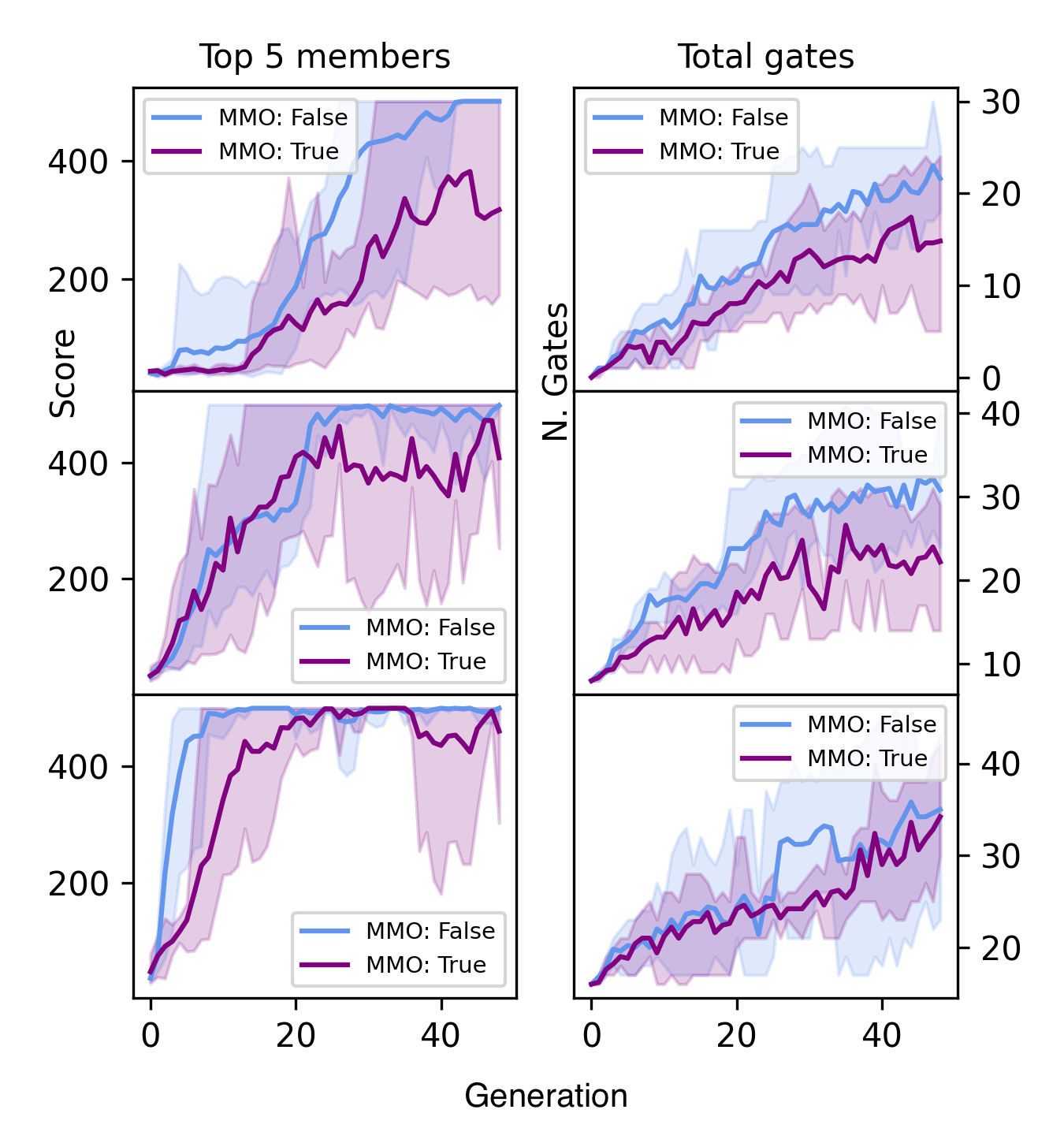}
        \caption{Comparison between QNEAT without the MOO alteration, and with the MOO. Each row represents an experiment performed with, respectively, 0, 1 and 2 initial layers, starting form the top row. In the first column, the average score of the top 5 members are shown, while in the right column the total gates are tracked. It can be seen how, as the number of initial layers increases, the averages of the top 5 members converge. Nonetheless, the number of gates is lower.}
        \label{fig:QNEAT-MOO}
    \end{figure}

    We show and compare the results of the QNEAT algorithm with and without the MOO variation. We evolve a VQC, both with and without MOO, starting from 0, 1 and 2 layers. The results are shown in Figure~\ref{fig:QNEAT-MOO}. It can be seen how the version with MOO performs worse both in terms of average and variance when it comes to the fitness of the environment. Nonetheless, it can be seen how the number of gates is actually lower. In the case of $2$ initial layers though, the difference between the two versions in not that sharp neither in the fitness nor in the number of gates, even though the QNEAT without any variation converges faster. 

    This behavior can be explained by the fact that the MOO algorithm looks for Pareto fronts in the space of fitnesses, where a front could be constituted of members that perform well but have a high number of gates, as well as members that have a small number of gates but perform poorly. Finding a good threshold to decide which part of the Pareto front to decide is essential to balance this possible trade-off.  

\section{Conclusions}
    In this paper, we explored the Quantum NeuroEvolution of Augmenting Topologies (QNEAT) algorithm. We defined how we adapted the genome and the process of crossover and mutation from the classical NEAT in order to be able to evolve the architectures of Variational Quantum Circuits. We then tested the algorithm of classical machine learning tasks, such as Reinforcement Learning and Combinatorial Optimization problems, and kept track during the evolution of the growth and depth of the architecture. In conclusion, we merged the QNEAT algorithm with the NSGA-2 algorithm for multi objective optimization. 
    
    One topic that should be studied in future works is the difference between the constrained and free architecture, in order to see how the last one would perform in terms of generations necessary to evolve a circuit up to the desired fitness, and also to see if an architecture can be found that will improve the score. Also, the QNEAT algorithm should be tested on harder and bigger problems, to see how effective it is on more complex reinforcement learning environments as well as bigger graphs for the case of the Max Cut problem.

\clearpage

\printbibliography

@article{ better1,
  author = "Aram W. Harrow and Ashley Montanaro",
  title = "Quantum computational supremacy",
  publisher = "Nature",
  year = "2017"
}

@article{ better2,  
  title = "Quantum Computation and Quantum Information",
  author = "Michael A. Nielsen, Isaac L. Chuang",
  publisher = "Cambridge",
  year = "2010"
}

@article{ better3,  
  title = "Polynomial-Time Algorithms for Prime Factorization and Discrete Logarithms on a Quantum Computer",
  author = "Peter W. Shor",
  publisher = "Siam",
  year = "1999"
}

@article{ universalapprox,
    title = "Approximation by superpositions of a sigmoidal function",
    author = "G. Cybenko",
    publisher = "Springen",
    year = "1989"
    
}

@article{ better4,  
  title = "Quantum Mechanics Helps in Searching for a Needle in a Haystack",
  author = "Lov K. Grover",
  publisher = "American Physical Society",
  year = "1997"
}

@article{ metrology1,  
  title = "Variational quantum algorithm for estimating the quantum Fisher information",
  author = "Jacob L. Beckey",
  publisher = "APS",
  year = "2022"
}

@article{ metrology2,  
  title = "Variational-state quantum metrology",
  author = "Bálint Koczor",
  publisher = "IOP",
  year = "2020"
}

@article{ math1,  
  title = "Variational Quantum Linear Solver",
  author = "Carlos Bravo-Prieto",
  year = "2019"
}

@article{ math2,  
  title = "Near-term quantum algorithms for linear systems of equations",
  author = "Hsin-Yuan Huang",
  year = "2019"
}

@article{ chem1,  
  title = "Quantum Chemistry in the Age of Quantum Computing",
  author = "Yudong Cao",
  year = "2019"
}

@article{ optim1,  
  title = "Quantum computing for energy systems optimization: Challenges and opportunities",
  author = "Akshay Ajagekar and Fengqi You",
  year = "2019"
}

@article{ optim2,  
  title = "Quantum Computing for Finance: State-of-the-Art and Future Prospects",
  author = "Daniel J. Egger",
  year = "2020"
}

@article{ optim3,  
  title = "Quantum Optimization and Quantum Learning: A Survey",
  author = "Yangyang Li",
  year = "2020"
}

@article{ hybrid1,  
  title = "Hybrid quantum-classical algorithms and quantum error mitigation",
  author = "Suguru Endo",
  year = "2020"
}

@article{ prob1,  
  title = "Towards fault-tolerant quantum computing with trapped ions",
  author = "Jan Benhelm",
  year = "2008"
}

@article{ prob2,  
  title = "Noise-induced barren plateaus in variational quantum algorithms",
  author = "Samson Wang",
  year = "2021"
}

@article{ qml,  
  title = "Quantum machine learning",
  author = "Jacob Biamonte",
  year = "2017"
}

@article{ bp1,  
  title = "Barren plateaus in quantum neural network training landscapes",
  author = "Jarrod R. McClean",
  year = "2018"
}

@article{ bp2,  
  title = "Cost function dependent barren plateaus in shallow parametrized quantum circuits",
  author = "M. Cerezo",
  year = "2021"
}

@article{ bp3,  
  title = "Absence of Barren Plateaus in Quantum Convolutional Neural Networks",
  author = "Arthur Pesah",
  year = "2021"
}

@article{ bp4,  
  title = "Barren Plateaus Preclude Learning Scramblers",
  author = "Zoë Holmes",
  year = "2021"
}

@article{ bp5,  
  title = "Analyzing the barren plateau phenomenon in training quantum neural networks with the ZX-calculus",
  author = "Chen Zhao",
  year = "2021"
}

@article{ bp6,  
  title = "Barren plateaus in quantum neural network training landscapes",
  author = "Jarrod R. McClean",
  year = "2018"
}

@article{ swarm,  
  title = "BParticle swarm optimization algorithm and its parameters: A review",
  author = "Mudita Juneja",
  year = "2016"
}

@article{ nes,  
  title = "Natural Evolution Strategies",
  author = "Daan Wierstra",
  year = "2014"
}

@article{ genetic,  
  title = "A review on genetic algorithm: past, present, and future",
  author = "Sourabh Katoch",
  year = "2020"
}

@article{ genetic2,  
  title = "Natural Evolutionary Strategies for Variational Quantum Computation",
  author = "Abhinav Anand",
  year = "2021"
}

@article{ qc1,  
  title = "Quantum Chemistry in the Age of Quantum Computing",
  author = "Yudong Cao",
  year = "2018"
}

@article{ qc2,  
  title = "Coupled-cluster theory in quantum chemistry",
  author = "Rodney J. Bartlett",
  year = "2007"
}

@article{ qaoa1,  
  title = "A Quantum Approximate Optimization Algorithm",
  author = "Edward Farhi",
  year = "2014"
}

@article{ qaoa2,  
  title = "From the Quantum Approximate Optimization Algorithm to a Quantum Alternating Operator Ansatz",
  author = "Stuart Hadfield",
  year = "2019"
}

@article{ hard-eff,  
  title = "Hardware-efficient variational quantum eigensolver for small molecules and quantum magnets",
  author = "Abhinav Kandala",
  year = "2017"
}

@article{ ADAPT-VQE,  
  title = "An adaptive variational algorithm for exact molecular simulations on a quantum computer",
  author = "Harper R. Grimsley",
  year = "2019"
}

@article{ EVQE,  
  title = "A Domain-agnostic, Noise-resistant, Hardware-efficient Evolutionary Variational Quantum Eigensolver",
  author = "Arthur G. Rattew",
  year = "2020"
}

@article{ tm1,  
  title = "Automated Design of Quantum Circuits",
  author = "Colin P. Williams",
  year = "1999"
}

@article{ tm2,  
  title = "Evolving Quantum Circuits using Genetic Algorithms",
  author = "Prashant",
  year = "2007"
}

@article{ tm3,  
  title = "Evolutionary Approach to Quantum and Reversible Circuits Synthesis",
  author = "Martin Lukac",
  year = "2003"
}

@article{ tm4,  
  title = "Evolving quantum circuits at the gate level with a hybrid quantum-inspired evolutionary algorithm",
  author = "Shengchao Ding",
  year = "2008"
}

@article{ tm5,  
  title = "Synthesis of reversible logic circuit using a species conservation method",
  author = "Xiaoxiao Wang",
  year = "2014"
}

@article{ neat,  
  title = "Evolving Neural Networks through Augmenting Topologies",
  author = "Kenneth O. Stanley",
  year = "2002"
}

@article{ qrl,  
  title = "A Survey on Quantum Reinforcement Learning",
  author = "Nico Meyer",
  year = "2022"
}

@article{ qrl2,  
  title = "Variational Quantum Circuits for Deep Reinforcement Learning",
  author = "Samuel Yen-Chi Chen",
  year = "2020"
}

@article{ qrl3,  
  title = "Reinforcement Learning with Quantum Variational Circuits",
  author = "Owen Lockwood",
  year = "2020"
}

@article{ qrl4,  
  title = "Quantum agents in the Gym: a variational quantum algorithm for deep Q-learning",
  author = "Andrea Skolik",
  year = "2022"
}

@article{ qveigen,  
  title = "The Variational Quantum Eigensolver: a review of methods and best practices",
  author = "Jules Tillya",
  year = "2022"
}

@article{ graphspaper,  
  title = "Learning to Optimize Variational Quantum Circuits to Solve Combinatorial Problems",
  author = "Sami Khairy",
  year = "2019"
}

@article{ nsga2,  
  title = "A fast and elitist multiobjective genetic algorithm: NSGA-II",
  author = "K. Deb",
  year = "2002"
}

\end{document}